\documentclass[10pt,oneside,final]{IEEEtran}
\usepackage{amssymb}
\usepackage{amsmath}

\usepackage{graphicx}
\usepackage{color}
\usepackage{multicol}

\usepackage{subfigure}
\usepackage{bm}
\usepackage{latexsym}
\usepackage{stfloats}
\usepackage{float}
\usepackage{exscale}
\usepackage{relsize}
\usepackage{cite}
\usepackage{cases}
\usepackage{algorithm}
\usepackage{algpseudocode}
\usepackage{amsthm}
\usepackage{epsfig}
\usepackage{epstopdf}
\usepackage{breqn}
\usepackage{enumerate}
\usepackage{multirow}
\usepackage[utf8]{inputenc}
\usepackage{algorithmicx}

\usepackage{diagbox} 
\usepackage{multirow} 

\begin{document}
\title{\vspace{-0.5cm}Deep Learning-Based CSI Feedback for RIS-Aided Massive MIMO Systems with Time Correlation}
\author{Zhangjie~Peng,
	    Zhaotian~Li,
	    Ruijing~Liu,
	    Cunhua~Pan,~\IEEEmembership{Senior Member,~IEEE},
	    
	    Feiniu~Yuan,~\IEEEmembership{Senior Member,~IEEE},
	    and Jiangzhou~Wang,~\IEEEmembership{Fellow,~IEEE}\vspace{-0.5cm}
\thanks{Z. Peng, Z. Li, R. Liu and F. Yuan are with the College of Information, Mechanical and Electrical Engineering, Shanghai Normal University, Shanghai 200234, China. (e-mail: pengzhangjie@shnu.edu.cn; 1000511820@smail.shnu.edu.cn; 1000511821@smail.shnu.edu.cn; yfn@shnu.edu.cn).}
\thanks{C. Pan is with the National Mobile Communications Research Laboratory, Southeast University, Nanjing 210096, China. (e-mail: cpan@seu.edu.cn).}
\thanks{J. Wang is with the School of Engineering, University of Kent, CT2 7NT Canterbury, U.K. (e-mail: j.z.wang@kent.ac.uk).}
}

\maketitle
\newtheorem{lemma}{Lemma}
\newtheorem{theorem}{Theorem}
\newtheorem{remark}{Remark}
\newtheorem{corollary}{Corollary}
\newtheorem{proposition}{Proposition}\vspace{-0.1cm}
\begin{abstract}
In this paper, we consider an reconfigurable intelligent surface (RIS)-aided frequency division duplex (FDD) massive multiple-input multiple-output (MIMO) downlink system. In the FDD systems, the downlink channel state information (CSI) should be sent to the base station through the feedback link. However, the overhead of CSI feedback occupies substantial uplink bandwidth resources in RIS-aided communication systems. In this work, we propose a deep learning (DL)-based scheme to reduce the overhead of CSI feedback by compressing the cascaded CSI. In the practical RIS-aided communication systems, the cascaded channel at the adjacent slots inevitably has time correlation. We use long short-term memory to learn time correlation, which can help the neural network to improve the recovery quality of the compressed CSI. Moreover, the attention mechanism is introduced to further improve the CSI recovery quality. Simulation results demonstrate that our proposed DL-based scheme can significantly outperform other DL-based methods in terms of the CSI recovery quality.
\end{abstract}
\begin{IEEEkeywords}
Time correlation, reconfigurable intelligent surface (RIS), CSI feedback, deep learning, attention mechanism.\vspace{-0.2cm}
\end{IEEEkeywords}
\section{Introduction}
Recently, of the key technologies of future sixth generation (6G) communication, the reconfigurable intelligent surface (RIS) has attracted extensive attention due to its low-cost and easy-deployment\cite{1,21}. Since RIS is an array of passive reflecting elements which can be deployed on the building surface to enhance signal transmission, it has been applied in the scenarios such as unmanned aerial vehicle communication, device-to-device communication, and mobile edge computing\cite{2,5,20,22}. However, the large number of reflective elements within the RIS leads to a high pilot overhead for the base station (BS) to acquire the cascaded channel state information (CSI) in RIS-aided systems. In order to acquire CSI, researchers have proposed methods such as CSI estimation and CSI feedback. Due to the channel reciprocity of time division duplex (TDD) systems, the BS can acquire the downlink CSI by estimating the uplink CSI. In RIS-aided wireless communication systems, researchers mainly focused on CSI estimation in TDD systems, while the RIS-aided wireless communication systems under the frequency division duplex (FDD) mode were ignored although it is the main operation model under low frequency band.

In FDD systems, due to the lack of channel reciprocity, the downlink CSI needs to be fed back from users to the BS through the feedback link. The overhead of the CSI feedback occupies substantial uplink bandwidth resources in RIS-aided systems, which affects the communication quality of the systems. Therefore, some contributions are proposed to reduce feedback overhead\cite{6,7,14}. In \cite{6}, a method based on codebook was proposed to compress CSI in massive multiple-input multiple-output (MIMO) systems. The author of \cite{7} proposed a method based on compressive sensing. In \cite{14}, researchers studied the RIS-aided massive MIMO system and proposed a CSI feedback method based on the codebook. However, these methods are not satisfactory in terms of the complexity of the algorithm and the accuracy of the decompression CSI.

In order to solve the problems that are difficult to accurately describe with mathematical models, deep learning (DL) has been introduced in wireless communications due to the success of DL in many areas in the past decade \cite{9,10,11,13,19,12}. In \cite{9}, researchers provided an overview of DL-based CSI feedback in massive MIMO systems. The author of \cite{10} first proposed a DL-based method to compress and decompress the CSI matrices which regarded CSI as an image, and designed the CsiNet referring to the structure of the automatic encoder applied in the image compression field. Based on CsiNet, the author of \cite{11} proposed channel reconstruction network (CRNet) which introduced the multi-resolution architecture to make the network perform better. In addition, the CSI at adjacent slots has the time correlation in the actual scenario. Thus the learning results of the previous CSI can assist in the learning of the next CSI. By considering the time correlation of the channels, a network was proposed based on long short-term memory (LSTM) \cite{13}. In \cite{19}, the authors proposed ConvlstmCsiNet based on convolutional LSTM (ConvLSTM) to achieve better performance by changing the weight calculation method. Currently, researches on DL-based CSI feedback in RIS-aided wireless communication systems is relatively limited, and the existing researches have not considered the time correlation of the channel.

The importance of the information obtained by neural networks is different during the learning process. If the network can distinguish and extract the information which is more important, the performance of the network can improve significantly. To solve this problem, the attention mechanism is used to distinguish the information which is more valuable through a small number of parameters. By introducing the attention mechanism, the performance of the network can be significantly improved while the complexity of the network has almost no increase. The author of \cite{12} proposed a network to enhance the CSI decompression performance in massive MIMO systems.

In this paper, we investigate the CSI feedback in the RIS-aided FDD massive MIMO systems. The contributions of this work are summarized as follows: 1) We present the time correlation model for the cascaded channel of the RIS-aided systems; 2) A network called attention convolution CsiNet (ACNet) is proposed to compress and decompress the CSI of the cascaded channel; 3) The numerical results demonstrate that the proposed network can significantly improve the performance of CSI decompression while the complexity of the network is basically unchanged.

\section{System Model}
\begin{figure}[t]
	\centering
	\vspace{-7pt}
	\includegraphics[width=0.8\linewidth]{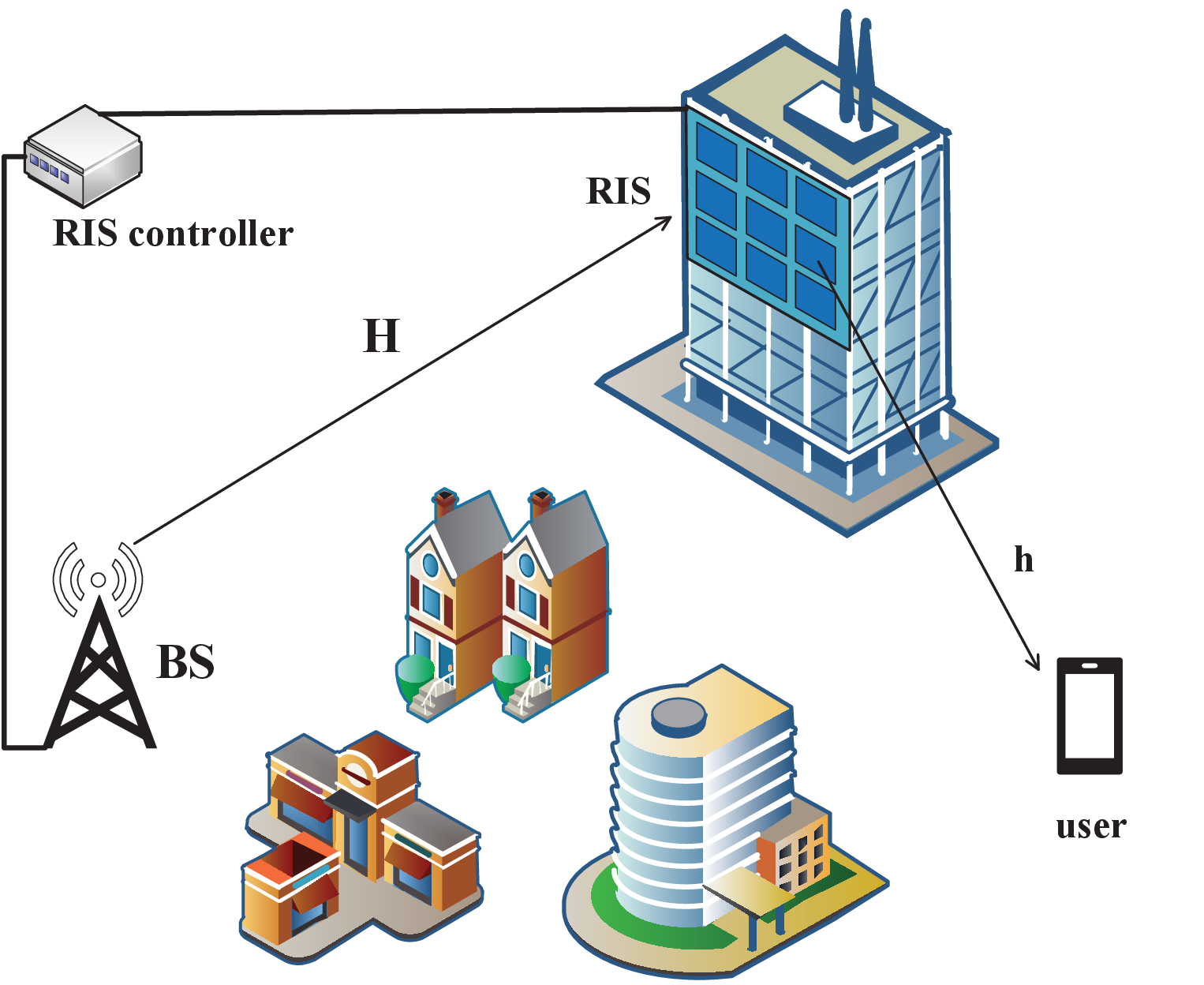}
	\caption{System model of RIS-aided massive MIMO system.} \label{fig1}\vspace{-0.4cm}
\end{figure}

As shown in Fig. \ref{fig1}, we consider an RIS-aided FDD massive MIMO downlink system where a user receives signals from the BS. The user is equipped with a single antenna, the BS is equipped with $M$ transmit antennas, and an RIS consisting of $N$ reflecting elements is deployed on the building surface. In addition, due to the existence of many obstacles, it is assumed that the direct channel between the BS and the user is blocked.\vspace{-0.2cm}

\subsection{Cascaded Channel Model}
We denote $\textbf{H}\in \mathbb{C}^{N\times M}$ as the channel matrix between the RIS and the BS, $\textbf{h}\in \mathbb{C}^{N\times 1}$ as the channel vector between the user and the RIS. $\textbf{H}$ and $\textbf{h}$ can be modeled as
\setlength\abovedisplayskip{2pt}
\setlength\belowdisplayskip{2pt}
\begin{align}\label{d}
\textbf{H}&=\sum_{i=1}^{L_{1}}\rho _{i}\textbf{m}\left ( p_{1,i},q_{1,i}\right )\textbf{n}^{H}\left ( p_{i}^{AOD}\right ),\\
\textbf{h}&=\sum_{i=1}^{L_{2}}\xi _{i}\textbf{m}\left ( p_{2,i},q_{2,i}\right ),
\end{align}
respectively, where $\rho _{i}$ and $\xi _{i}$ denote the gain of the $i$-th path, $L_{1}$ and $L_{2}$ are the numbers of paths of the BS-RIS channel and RIS-user channel, respectively. In addition, steering vector $\textbf{n}\left ( p_{i}^{AOD}\right )$ denotes the antenna array response of the $i$-th path, $\textbf{m}\left ( p_{1,i},q_{1,i}\right )$ is the steering vector of the BS-RIS channel of the $i$-th path, and $\textbf{m}\left ( p_{2,i},q_{2,i}\right )$ is the steering vector of the RIS-user channel of the $i$-th path. $\textbf{m}\left ( p_{1,i},q_{1,i}\right )$, $\textbf{n}\left ( p_{i}^{AOD}\right )$, and $\textbf{m}\left (p_{2,i},q_{2,i}\right )$ are expressed as
\setlength\abovedisplayskip{2pt}
\setlength\belowdisplayskip{2pt}
\begin{align}\label{e}
\!\!&\textbf{m}\left ( p_{1,i},q_{1,i}\right )\!=\!\frac{1}{\sqrt{N_{1}}}\!\left [ e^{j2\pi n_{1}p_{1,i}} \right ]^{T}\!\otimes\!\frac{1}{\sqrt{N_{2}}}\left [ e^{j2\pi n_{2}q_{1,i}} \right ]^{T}\!,\\
\!\!&\textbf{n}\left ( p_{i}^{AOD}\right )=\frac{1}{\sqrt{M}}\!\left [ e^{j2\pi mp_{i}^{AOD}} \right ]^{T},\\
\!\!&\textbf{m}\left ( p_{2,i},q_{2,i}\right )\!=\!\frac{1}{\sqrt{N_{1}}}\!\left [ e^{j2\pi n_{1}p_{2,i}} \right ]^{T}\!\otimes\!\frac{1}{\sqrt{N_{2}}}\left [ e^{j2\pi n_{2}q_{2,i}} \right ]^{T}\!,
\end{align}
respectively, where $\otimes$ denotes the Kronecker product. In addition, $n_{1}\in \left\{ 1,2,\cdots,N_{1}\right\}$,  $n_{2}\in \left\{ 1,2,\cdots,N_{2}\right\}$, and $m\in \left\{ 1,2,\cdots,M\right\}$, where $N_{1}$ and $N_{2}$ represent the numbers of elements in the horizontal and the vertical directions of the RIS, respectively. $p_{1,i}=\frac{d_{1}}{\lambda }cos\beta _{BR,i}sin\alpha _{BR,i}$ and $q_{1,i}=\frac{d_{1}}{\lambda }sin\beta _{BR,i}$ denote the normalized spacial azimuth angles of arrival (AoAs) and the normalized spacial elevation AoAs for the RIS with the range of $\left [ -\frac{1}{2},\frac{1}{2} \right ]$, respectively. $p_{2,i}=\frac{d_{1}}{\lambda }cos\beta _{RU,i}sin\alpha _{RU,i}$ and $q_{2,i}=\frac{d_{1}}{\lambda }sin\beta _{RU,i}$ denote the normalized spacial azimuth angles of departure (AoDs) and the normalized spacial elevation AoDs for the RIS with the range of $\left [ -\frac{1}{2},\frac{1}{2} \right ]$, respectively. $p_{i}^{AOD}=\frac{d_{2}}{\lambda }sin\alpha_{i}^{AOD}$ denotes the normalized spacial AoD for the BS with the range of $\left [ -\frac{1}{2},\frac{1}{2} \right ]$. $\lambda$ represents the wavelength, $d_{1}$ and $d_{2}$ denote the element spacing at the RIS and the antenna spacing at the BS, respectively, and we assume $d_{1}=d_{2}=\lambda/2$. 

In this work, we assume that the CSI of the cascaded channel is perfect. Therefore, we focus on the compression and decompression of the CSI matrix. Then, the signal received by the user is given by
\setlength\abovedisplayskip{2pt}
\setlength\belowdisplayskip{2pt}
\begin{align}\label{eqn:2}
y= \textbf{h}_{}^{H}\textrm{diag}(\textbf{e})\textbf{H}\sqrt{p}\textbf{v}x+n,
\end{align}
where $\textbf{e}\in \mathbb{C}^{N\times 1}$ denotes the phase shift vector of the RIS, $x$ denotes the transmitted signal of the BS, $n$ denotes the additive white Gaussian noise, $p$ is the transmit power of the BS, and $\textbf{v}\in \mathbb{C}^{M\times 1}$ denotes the precoding vector. By utilizing the property of diagonal matrix, the expression (\ref{eqn:2}) can be reformulated as
\setlength\abovedisplayskip{2pt}
\setlength\belowdisplayskip{2pt}
\begin{equation}\label{eqn:1}
\begin{aligned}
y\!&=\textbf{e}_{}^{T}\textrm{diag}(\textbf{h}_{}^{H})\textbf{H}\sqrt{p}\textbf{v}x+n\\
 \!&=\textbf{e}_{}^{T}\textbf{G}\sqrt{p}\textbf{v}x+n,
\end{aligned}
\end{equation}
where $\textbf{G}=\textrm{diag}(\textbf{h}_{}^{H})\textbf{H}$ denotes the cascaded channel matrix of the RIS-aided system. \vspace{-0.2cm}

\subsection{Time Correlation Model}
In the actual scenario, the location of the user will change over time, and the environment around the user does not fully change. In a short time, e.g., feedback interval, the moving distance of the user is small. For example, even if the moving speed of the user reaches $\rm{360\,km/h}$, the moving distance of the user within one millisecond is only $\rm{0.1\,m}$. Therefore, the propagation environment at adjacent slots is very similar. As the CSI is determined by the propagation environment, the CSI at adjacent slots exhibits a high correlation. By using the first-order Markov process in \cite{19}, the time change of CSI can be expressed as
\setlength\abovedisplayskip{2pt}
\setlength\belowdisplayskip{2pt}
\begin{align}\label{8}
\textbf{G}_{t}=\beta \textbf{G}_{t-1}+\gamma \textbf{U}_{t},
\end{align}
where $\textbf{G}_{t}$ denotes the CSI matrix of cascaded channel at the $t$-th time step. We set related parameter $\alpha\in (0,1]$. The time correlation coefficient is $ \beta = 1-\alpha^{2}$ and the noise correlation coefficient is $\gamma = \alpha^{2}$. It is obvious that $\alpha = 1$ indicates that the CSI at adjacent slots does not correlate, while it can generate a time-invariant CSI matrix when $\alpha\to 0$. $\textbf{U}_{t}\in \mathbb{C}^{N\times M}$ is the additive noise that each element follows the distribution of $u_{t} \sim N\left (0, \sigma ^{2} \right)$. The sequence of the channel matrices can be defined as $\left\{ \textbf{G}\right\}_{t=1}^{T}=\left\{ \textbf{G}_{1},\textbf{G}_{2},...,\textbf{G}_{T}\right\}$.

\begin{figure*}[htbp]	
	\centering	
	\includegraphics[height=\linewidth,angle=90]{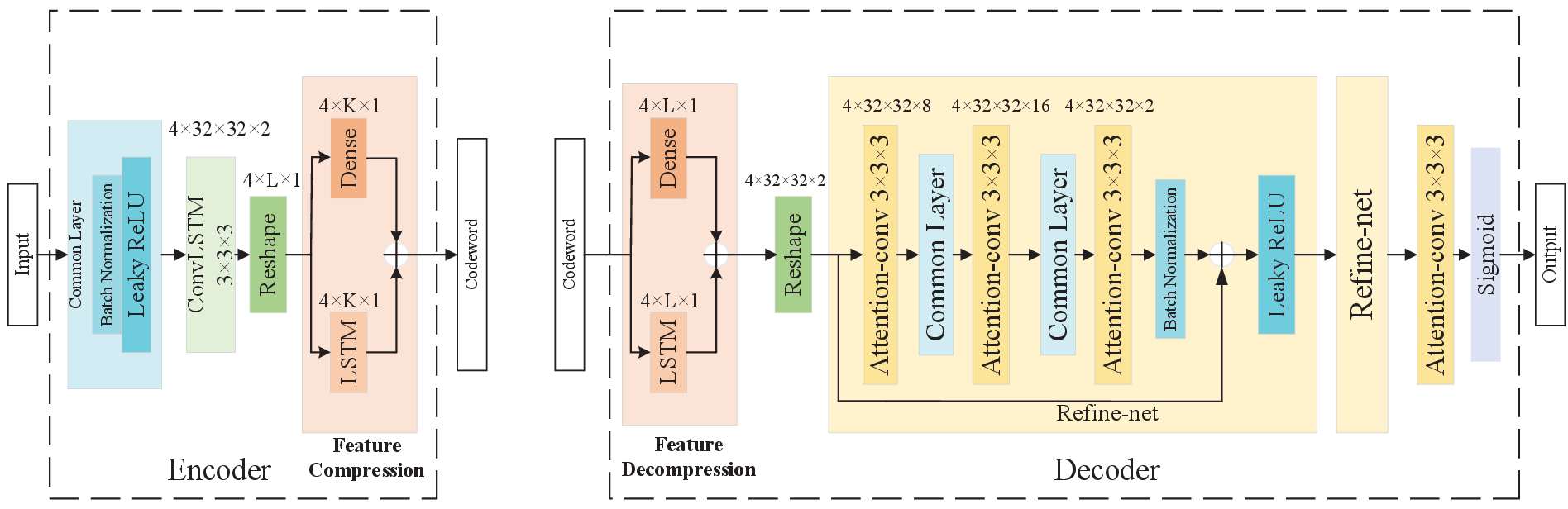}	
	\caption{The architecture of ACNet.}	
	\label{FigureTwo}\vspace{-0.4cm}
\end{figure*}
\section{CSI Feedback Process And ACNet}
In this section, we will introduce the process of CSI feedback and the architecture of ACNet. The network layer and the differences in network design will be introduced in details.\vspace{-0.3cm}
\subsection{CSI feedback process}
Due to the incapability of neural networks to handle complex numbers, $\textbf{G}$ is divided into a real part and an imaginary part, and the size of the CSI matrices is $L=2NM$. Denote $\textbf{G}_{\textrm{in}}$ as the divided CSI matrices. All elements in the $\textbf{G}_{\textrm{in}}$ are normalized within $[0,1]$. The encoder of the network will compress the $L$ size CSI matrix $\textbf{G}_{\textrm{in}}$ into a $K$-dimensional feature vector $\textbf{s}$ based on the given compression ratio ($C\!R$), which can be expressed as
\setlength\abovedisplayskip{2pt}
\setlength\belowdisplayskip{2pt}
\begin{align}\label{a}
	C\!R=\frac{L}{K},
\end{align}
and the process of compressing can be expressed as
\setlength\abovedisplayskip{2pt}
\setlength\belowdisplayskip{2pt}
\begin{align}\label{10}
&\textbf{s}=f_{\textrm{en}}\left ( \textbf{G}_{\textrm{in}},\theta _{\textrm{en}} \right ),
\end{align}
where $f_{\textrm{en}}\left ( \cdot  \right )$ represents the compression function and $\theta _{\textrm{en}}$ denotes the parameters of the encoder. 

When the feature vector $\textbf{s}$ is received by the BS, the decoder of the network will decompress the $K$-dimensional feature vector $\textbf{s}$ into the $L$ size CSI matrix $\textbf{G}_{\textrm{out}}$. The process of decompressing can be expressed as
\setlength\abovedisplayskip{2pt}
\setlength\belowdisplayskip{2pt}
\begin{align}\label{11}
&\textbf{G}_{\textrm{out}}=f_{\textrm{de}}\left ( \textbf{s},\theta _{\textrm{de}} \right ),
\end{align}
where $f_{\textrm{de}}\left ( \cdot  \right )$ represents the decompression function and $\theta _{\textrm{de}}$ denotes the parameters of the encoder.
The loss function of ACNet is set to mean square error (MSE). By substituting (\ref{10}) and (\ref{11}) into the MSE function, the optimized expression of ACNet is given by
\setlength\abovedisplayskip{2pt}
\setlength\belowdisplayskip{2pt}
\begin{align}\label{12}
&\left ( {\hat{\theta}}_{\textrm{en}},{\hat{\theta}}_{\textrm{de}} \right )=\mathop{\arg\min\limits_{\theta _{\textrm{en}},\theta _{\textrm{de}}}}\left\| \textbf{G}_{\textrm{out}}-\textbf{G}_{\textrm{in}}\right\|_{2}^{2},
\end{align}
where ${\hat{\theta}}_{\textrm{en}}$ and ${\hat{\theta}}_{\textrm{de}}$ denote the optimal parameters of $f_{\textrm{de}}\left ( \cdot  \right )$ and $f_{\textrm{en}}\left ( \cdot  \right )$.
\begin{figure}[t]
	\centering
	\includegraphics[width=\linewidth]{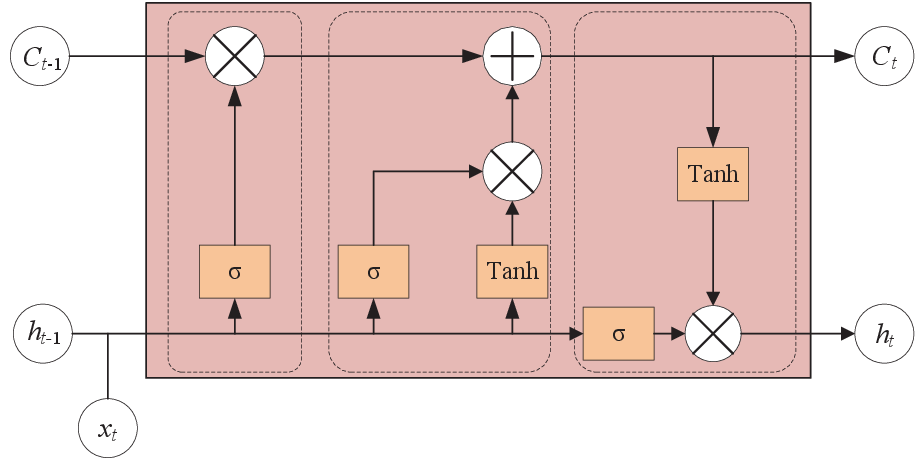}
	\caption{The structures of ConvLSTM and LSTM.} \label{FigureLSTM}\vspace{-0.4cm}
\end{figure}
\subsection{Architecture of ACNet}
\subsubsection{ACNet}
The main structure of ACNet is shown in Fig. \ref{FigureTwo}. Different network layers are marked with different colors. The output of each layer is marked on the top. On the basis of the ConvlstmCsiNet, we improve the performance of Refine-net by adding the attention mechanism.

The encoder is composed of a ConvLSTM module and a feature compression module. The ConvLSTM converts the weight calculation method from linear operation to convolution operation on the basis of the LSTM\cite{16}. It cannot only obtain the time correlation in the sequence data, but also describe local details of the image like the convolutional neural network.

Fig. \ref{FigureLSTM} is the structure of the ConvLSTM and the LSTM. As mentioned above, they are the same in the structure, but different in the weight calculation method. $x_{t}$, $h_{t}$, and $C_{t}$ denote the input, the output, and the current state at the $t$-th time step, respectively. The LSTM is composed of three gates, which are the forget gate, the input gate, and the output gate. These three gates can process $x_{t}$ and $h_{t-1}$, and add or delete information to the cell state according to the processing results. The forget gate determines whether to discard the previous state information $C_{t-1}$ based on the input data. The function of the input gate is to decide whether $x_{t}$ and $h_{t-1}$ need to be stored in the cell and discard the unwanted information. Finally, the output gate outputs $h_{t}$ which is decided by $x_{t}$, $h_{t-1}$, and the current state $C_{t}$.

The feature compression and decompression modules are symmetric. The feature compression module is used to compress the reshaped $L$-length vector into a $K$-length codeword. And the feature decompression module is used to decompress the $K$-length codeword into an $L$-length vector. These modules consist of two parallel layers, i.e., the fully-connected layer and the LSTM layer\cite{17}. In the feature compression and decompression modules, we use the LSTM instead of the ConvLSTM. Due to its superior performance in overall information interaction, LSTM is better suited for feature compression, while ConvLSTM is more suitable for describing local details.

After decompression, the $L$-length vector will be reshaped into two $32\times 32$ sized matrices. Then, we use two RefineNet blocks, a $3\times 3\times 3$ Attention-conv layer, and a Sigmoid activation layer to recover the CSI matrices.
\begin{figure}[t]
	\centering
	\includegraphics[width=7cm]{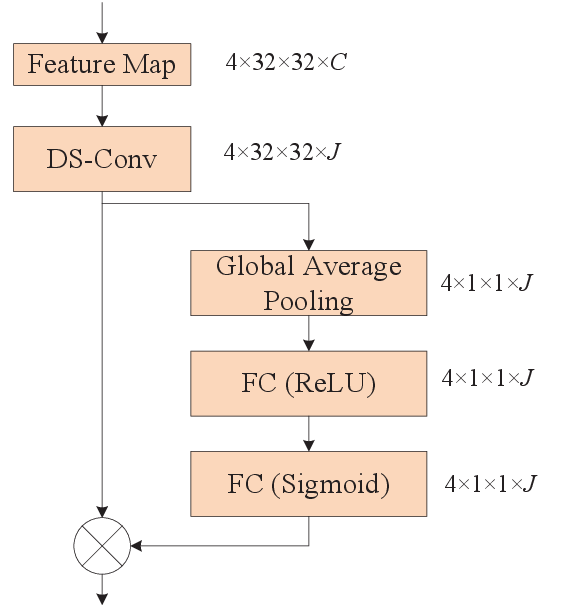}
	\caption{The structure of Attention-conv.} \label{FigureThree}\vspace{-0.4cm}
\end{figure}
\subsubsection{Attention Mechanism}
The attention mechanism can help the network extract more valuable information to achieve better performance while the complexity of the network has almost no increase \cite{21}. We introduce the attention mechanism to the network by adding Attention-conv. Fig. \ref{FigureThree} shows the structure of Attention-conv.

When we get $4\times 32\times 32\times C$ feature maps, we first use the depthwise separable convolution (DS-Conv) to get $4\times 32\times 32\times J$ feature maps \cite{18}. We use the DS-Conv instead of the Conv3D due to the fact that the DS-Conv not only has smaller number of parameters, but also can achieve a better CSI decompression performance than the Conv3D.

Then, we use a global average pooling to get a $4\times 1\times 1\times J$ vector. Next, we input the vector into two fully connected layers. The activation functions of two fully connected layers are ReLU and Sigmoid, respectively. Finally, we use this $J$-dimension vector to multiply the $4\times 32\times 32\times J$ feature map to get the final feature maps.

For each CSI matrix, the proportion of useful information on different feature maps is different. The information on some maps is significant for CSI matrix decompression, while the information on other maps is unimportant. By adding the attention mechanism, each CSI matrix can get different attention weights on different feature maps. With this strategy, the network can extract more valuable information, and the network can decompress the CSI matrix better. 

The network complexity analysis is shown in Table I. We use the number of parameters to express the complexity of the networks. In Table I, the CsiNet has the least number of parameters due to the simplest structure. Compared with the CsiNet, the complexity of the ConvlstmCsiNet significantly increases, and its complexity enhancement is mainly caused by the LSTM. Due to the advantage of the attention mechanism, the complexity of the ACNet has almost no increase compared with the ConvlstmCsiNet.

\begin{table}[t]
	\begin{center}
		\caption{The number of parameters under different $C\!R$ }
		\renewcommand\arraystretch{1.5}
		\setlength{\tabcolsep}{1.5mm}{
		\begin{tabular}{|c|c|c|c|c|} 
			\hline
			$C\!R$ & 4 & 8 & 16 & 32\\
			\hline
			CsiNet & 2,103,904 & 1,055,072 & 530,656 & 268,448\\
			\hline
			ConvlstmCsiNet & 28,326,896 & 22,296,304 & 19,477,616 & 18,117,424\\
			\hline
			\textbf{ACNet} & $\textbf{28,328,012}$  & $\textbf{22,297,420}$ & $\textbf{19,478,732}$ & $\textbf{18,118,540}$\\
			\hline
		\end{tabular}\vspace{-0.4cm}}
	\end{center}
\end{table}

\section{Numerical Results}
In this section, we present the simulation results of the CsiNet, the ConvlstmCsiNet, and the ACNet. We set $M=N=32$, $L_{1}=L_{2}=3$, $N_{1}=4$, and $N_{2}=8$. In the network training process, we set the epoch to 100, the batch size to 250, and the learning rate to 0.001. The number of samples in the training set, the test set, and the verification set are 25000, 5000, and 5000.

From (\ref{8}), we add related parameter $\alpha$ and a tiny white Gauss noise ($\sigma=10^{-3}$) between each time step. And we can extend the two-dimensional CSI feedback matrix to the $T$-time sequence of the time-varying CSI matrices. In this work, we set $T = 4$ for convenience\cite{19}.

Cosine similarity ($\rho$) and normalized mean square error (NMSE) are introduced to compare the performance of different networks\cite{19}. $\rho$ can be expressed as
\setlength\abovedisplayskip{2pt}
\setlength\belowdisplayskip{2pt}
\begin{align}\label{gtiimp}
&\rho =\mathbb{E}\left\{ \frac{1}{T}\frac{1}{N_{c}}\sum_{t=1}^{T}\sum_{n=1}^{N_{c}}\frac{\left|\textbf{g}_{\textrm{out}\left ( n,t \right )}^{H}\cdot \textbf{g}_{\textrm{in}\left ( n,t \right )} \right|}{\left\| \textbf{g}_{\textrm{out}\left ( n,t \right )}\right\|_{2}\left\| \textbf{g}_{\textrm{in}\left ( n,t \right )}\right\|_{2}}\right\},
\end{align}
where $N_{c}$ represents the number of columns of the matrix, $\textbf{g}_{\textrm{out}\left ( n,t \right )}$ and $\textbf{g}_{\textrm{in}\left ( n,t \right )}$ represent the column vector of the output matrix and the input matrix, respectively. And the NMSE can be expressed as
\setlength\abovedisplayskip{2pt}
\setlength\belowdisplayskip{2pt}
\begin{align}\label{gtiimp}
&\textrm{NMSE}=\textrm{log}\left (\mathbb{E}\left\{ \frac{1}{T}\sum_{t=1}^{T}\frac{\left\| \textbf{G}_{\textrm{in}}-\textbf{G}_{\textrm{out}} \right\|_{F}^{2}}{\left\| \textbf{G}_{\textrm{in}}\right\|_{F}^{2}}\right\}\right ).
\end{align}

The performance of networks is given in Table II. The best performance under the same $C\!R$ is shown in bold. Obviously, the proposed ACNet achieves the best performance under different $C\!R$.

\begin{table}[t]
	\begin{center}
		\caption{The performance of different net at $\alpha=0.1$}
		\renewcommand\arraystretch{1.2}
		\begin{tabular}{|c|c|c|c|c|c|} 
			\hline
			& $C\!R$ & 4 & 8 & 16 & 32\\
			\hline
			\multirow{3}{0.1cm}{\rotatebox{90}{NMSE}} & CsiNet & -20.46 & -13.57 & -7.55 & -4.73\\
			\cline{2-6}
			& ConvlstmCsiNet & -19.52 & -17.04 & -14.57 & -10.20\\
			\cline{2-6}
			& \textbf{ACNet} & $\textbf{-22.74}$ & $\textbf{-19.46}$ & $\textbf{-17.39}$ & $\textbf{-12.86}$\\
			\hline
			\multirow{3}{0.1cm}{$\rho$} & CsiNet & 98.05$\%$ & 93.39$\%$ & 81.25$\%$ & 73.76$\%$\\
			\cline{2-6}
			& ConvlstmCsiNet & 97.04$\%$ & 95.21$\%$ & 93.98$\%$ & 87.60$\%$\\
			\cline{2-6}
			& \textbf{ACNet} & $\textbf{98.69\%}$ & $\textbf{96.85\%}$ & $\textbf{96.86\%}$ & $\textbf{91.81\%}$\\
			\hline
		\end{tabular}\vspace{-0.3cm}
	\end{center}
	
\end{table}
\begin{table}[htbp]
	\begin{center}
		\caption{The percentage improvement about NMSE of the proposed network}
		\renewcommand\arraystretch{1.5}
		\setlength{\tabcolsep}{1.2mm}{
			\begin{tabular}{|c|c|c|c|c|} 
				\hline
				$C\!R$ & 4 & 8 & 16 & 32\\
				\hline
				Compare to CsiNet & 11.1$\%$ & 43.4$\%$ & 130.3$\%$ & 171.9$\%$\\
				\hline
				Compare to ConvlstmCsiNet & 16.5$\%$ & 14.2$\%$ & 19.4$\%$ & 26.1$\%$\\
				\hline
				
			\end{tabular}\vspace{-0.5cm}}
	\end{center}
\end{table}

In order to show the comparison more clearly, we give the percentage of performance improvement of the ACNet compared with the ConvlstmCsiNet and the CsiNet in Table III. As shown in Table III, the performance improvement of the ACNet increases significantly compared with the CsiNet when the CR is high, which is due to the LSTM modules and the ConvLSTM modules. These modules can help the network learn the time correlation of the cascaded CSI. Compared with the ConvlstmCsiNet, the improvement of NMSE is about $\rm{15\,\%}$ when $C\!R$ equals 4, 8, and 16. When $C\!R = 32$, the improvement of NMSE is $\rm{26.1\,\%}$, which shows that the attention mechanism can achieve better results, especially for the high $C\!R$. This is because the network needs to obtain valuable information as much as possible to obtain better compression and decompression performance when the $C\!R$ is high. And the attention mechanism can help the network distinguish valuable information.

\begin{figure}[t]
	\centering
	\includegraphics[width=\linewidth]{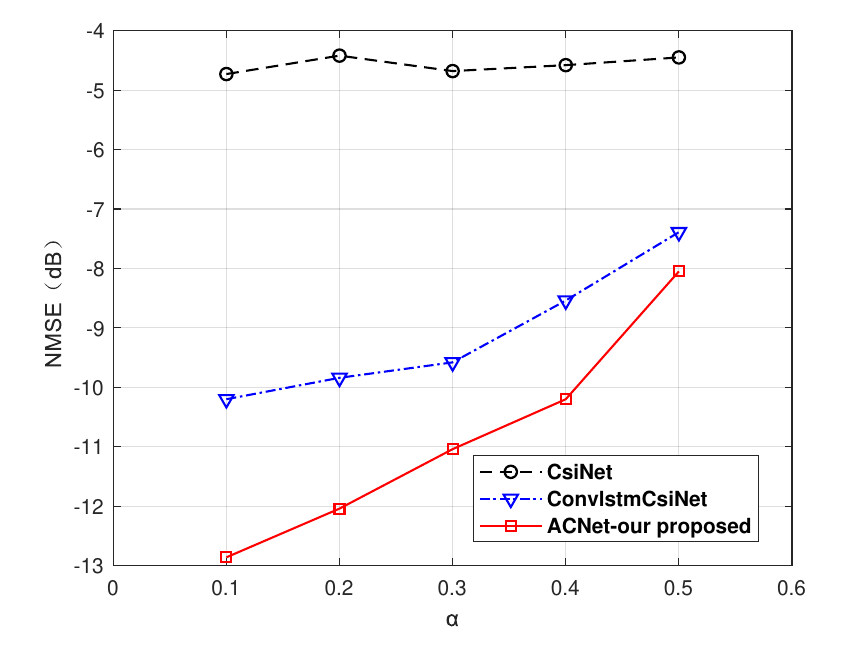}
	\caption{The corresponding NMSE of the networks under the different $\alpha $ and $C\!R=32$.} \label{f6}\vspace{-0.4cm}
\end{figure}

Fig. \ref{f6} shows the corresponding NMSE of the networks versus the values of $\alpha $ when $C\!R=32$. For CsiNet, changes in $\alpha $ will not significantly affect the performance of the network. On the contrary, the increase of $\alpha $ will lead to a growth of the corresponding NMSE for ConvlstmCsiNet and ACRNet. In addition, it can be seen that the performance of ACNet is better than ConvlstmCsiNet for the same $\alpha $. However, when $\alpha $ exceeds $\rm{0.4}$, the performance of ACNet significantly decreases due to the influence of noise. Therefore, ACNet is more suitable for scenarios with a high time correlation.\vspace{-0.3cm}

\section{Conclusion}
In this work, we proposed a DL-based network which can compress the CSI and reduce the overhead of CSI feedback. We used the LSTM to learn the time correlation of channels on adjacent slots and the attention mechanism has been added to improve compression performance. The results of numerical simulation demonstrated that the ACNet has a significant improvement under the different $C\!R$ compared with the ConvlstmCsiNet while the complexity of the ACNet was nearly unchanged compared with the ConvlstmCsiNet.

\bibliographystyle{IEEEtran}
\bibliography{IEEEabrv,Refer}

\begin{thebibliography}{10}
\providecommand{\url}[1]{#1}
\csname url@samestyle\endcsname
\providecommand{\newblock}{\relax}
\providecommand{\bibinfo}[2]{#2}
\providecommand{\BIBentrySTDinterwordspacing}{\spaceskip=0pt\relax}
\providecommand{\BIBentryALTinterwordstretchfactor}{4}
\providecommand{\BIBentryALTinterwordspacing}{\spaceskip=\fontdimen2\font plus
\BIBentryALTinterwordstretchfactor\fontdimen3\font minus
  \fontdimen4\font\relax}
\providecommand{\BIBforeignlanguage}[2]{{%
\expandafter\ifx\csname l@#1\endcsname\relax
\typeout{** WARNING: IEEEtran.bst: No hyphenation pattern has been}%
\typeout{** loaded for the language `#1'. Using the pattern for}%
\typeout{** the default language instead.}%
\else
\language=\csname l@#1\endcsname
\fi
#2}}
\providecommand{\BIBdecl}{\relax}
\BIBdecl

\bibitem{1}
Z.~Zhang, Y.~Xiao, Z.~Ma, M.~Xiao, Z.~Ding, X.~Lei, G.~K. Karagiannidis, and
  P.~Fan, ``6{G} wireless networks: Vision, requirements, architecture, and key
  technologies,'' \emph{IEEE Veh. Technol. Mag.}, vol.~14, no.~3, pp. 28--41,
  Sep. 2019.

\bibitem{21}
C.~Pan, G.~Zhou, K.~Zhi, S.~Hong, T.~Wu, Y.~Pan, H.~Ren, M.~Di~Renzo, A.~L.
  Swindlehurst, R.~Zhang \emph{et~al.}, ``An overview of signal processing
  techniques for ris/irs-aided wireless systems,'' \emph{IEEE J. Sel. Topics
  Signal Process.}, 2022.

\bibitem{2}
Z.~Peng, T.~Li, C.~Pan, H.~Ren, and J.~Wang, ``{RIS}-aided {D2D} communications
  relying on statistical {CSI} with imperfect hardware,'' \emph{IEEE Wireless
  Commun. Lett.}, vol.~26, no.~2, pp. 473--477, Dec. 2022.

\bibitem{5}
H.~Ren, Z.~Zhang, Z.~Peng, L.~Li, and C.~Pan, ``Energy minimization in
  {RIS}-assisted {UAV}-enabled wireless power transfer systems,'' \emph{IEEE
  Internet Things J.}, vol.~10, no.~7, pp. 5794--5809, Apr. 2023.

\bibitem{20}
Z.~Peng, R.~Weng, Z.~Zhang, C.~Pan, and J.~Wang, ``Active reconfigurable
  intelligent surface for mobile edge computing,'' \emph{IEEE Wireless Commun.
  Lett.}, vol.~11, no.~12, pp. 2482--2486, Sep. 2022.

\bibitem{22}
K.~Zhi, C.~Pan, H.~Ren, K.~K. Chai, and M.~Elkashlan, ``Active ris versus
  passive ris: Which is superior with the same power budget?'' \emph{IEEE
  Commun. Lett.}, vol.~26, no.~5, pp. 1150--1154, 2022.

\bibitem{6}
D.~J. Love, R.~W. Heath, V.~K. N.~Lau, D.~Gesbert, B.~D. Rao, and M.~Andrews,
  ``An overview of limited feedback in wireless communication systems,''
  \emph{IEEE J. Select. Areas Commun.}, vol.~26, no.~8, pp. 1341--1365, Oct.
  2008.

\bibitem{7}
Z.~Qin, J.~Fan, Y.~Liu, Y.~Gao, and G.~Y. Li, ``Sparse representation for
  wireless communications: A compressive sensing approach,'' \emph{IEEE Signal.
  Proc. Mag.}, vol.~35, no.~3, pp. 40--58, May. 2018.

\bibitem{14}
D.~Shen and L.~Dai, ``Dimension reduced channel feedback for reconfigurable
  intelligent surface aided wireless communications,'' \emph{IEEE Trans.
  Commun.}, vol.~69, no.~11, pp. 7748--7760, Nov. 2021.

\bibitem{9}
J.~Guo, C.-K. Wen, S.~Jin, and G.~Y. Li, ``Overview of deep learning-based
  {CSI} feedback in massive {MIMO} systems,'' \emph{IEEE Trans. Commun.},
  vol.~70, no.~12, pp. 8017--8045, Oct. 2022.

\bibitem{10}
J.~Li, Q.~Zhang, X.~Xin, Y.~Tao, Q.~Tian, F.~Tian, D.~Chen, Y.~Shen, G.~Cao,
  Z.~Gao, and J.~Qian, ``Deep learning-based massive {MIMO CSI} feedback,'' in
  \emph{2019 18th International Conference on Optical Communications and
  Networks (ICOCN)}, 2019, pp. 1--3.

\bibitem{11}
Z.~Lu, J.~Wang, and J.~Song, ``Multi-resolution {CSI} feedback with deep
  learning in massive {MIMO} system,'' in \emph{Proc. IEEE Int. Conf. Commun.
  (ICC)}, 2020, pp. 1--6.

\bibitem{13}
T.~Wang, C.-K. Wen, S.~Jin, and G.~Y. Li, ``Deep learning-based {CSI} feedback
  approach for time-varying massive {MIMO} channels,'' \emph{IEEE Wireless
  Commun. Lett.}, vol.~8, no.~2, pp. 416--419, Apr. 2019.

\bibitem{19}
X.~Li and H.~Wu, ``Spatio-temporal representation with deep neural recurrent
  network in {MIMO CSI} feedback,'' \emph{IEEE Wireless Commun. Lett.}, vol.~9,
  no.~5, pp. 653--657, May. 2020.

\bibitem{12}
Q.~Cai, C.~Dong, and K.~Niu, ``Attention model for massive {MIMO CSI}
  compression feedback and recovery,'' in \emph{Proc. IEEE Wireless Commun.
  Netw. Conf. (WCNC)}, 2019, pp. 1--5.

\bibitem{16}
X.~Shi, Z.~Chen, H.~Wang, D.-Y. Yeung, W.-K. Wong, and W.-c. Woo,
  ``Convolutional {LSTM} network: A machine learning approach for precipitation
  nowcasting,'' \emph{Advances in neural information processing systems}, 2015.

\bibitem{17}
C.~Lu, W.~Xu, H.~Shen, J.~Zhu, and K.~Wang, ``{MIMO} channel information
  feedback using deep recurrent network,'' \emph{IEEE Commun. Lett.}, vol.~23,
  no.~1, pp. 188--191, Jan. 2019.

\bibitem{18}
M.~Sandler, A.~Howard, M.~Zhu, A.~Zhmoginov, and L.-C. Chen, ``Mobilenetv2:
  Inverted residuals and linear bottlenecks,'' in \emph{2018 IEEE/CVF
  Conference on Computer Vision and Pattern Recognition}, Jun. 2018, pp.
  4510--4520.

\end{thebibliography}

\end{document}